\documentclass[a4paper,11pt]{article}
\usepackage{pos}
\usepackage{overpic}
\usepackage{booktabs} 
\usepackage{transparent}

\title{A model-driven search for extreme BL Lacs among LAT BCU}

\manuallySeparateAuthors
\author*[a,b,c]{M. Nievas Rosillo,}
\author[d,e]{G. Chiaro,}
\author[f]{A. Dominguez}
\author{and G. La Mura$^g$ on behalf of the Fermi Large Area Telescope Collaboration}

\affiliation[a]{Instituto de Astrof\'isica de Canarias (IAC) \\ C/Via Lactea S/N E-38205 La Laguna, Tenerife, Spain}

\affiliation[b]{Universidad de La Laguna, Dept. Astrof\'isica (ULL) \\
Av. Astrofisico Francisco Sánchez, S/N, E-38206 La Laguna, Tenerife, Spain}

\affiliation[c]{Deutsches Elektronen-Synchrotron (DESY) \\ Platanenallee 6, Zeuthen, Germany, Spain}

\affiliation[d]{Institute of Space Astrophysics and Cosmic Physics IASF / INAF  \\ Via A.Corti 12, I-20133 Milano Italy}

\affiliation[e]{Consorzio Interuniversitario per la Fisica Spaziale  CIFS \\ Via Pietro Giuria, 1, 10125 Torino IT}

\affiliation[f]{IPARCOS and Department of EMFTEL, Universidad Complutense de Madrid, E-28040 Madrid, Spain}

\affiliation[g]{Lab. de Instrumenta\c{c}\~ao e F\'{i}sica Experimental de Part\'{i}culas. LIP, \\ Av. Prof. Gama Pinto 2, 1649-003 Lisboa, Portugal}

\emailAdd{mnievas@iac.es}

\abstract{The emission of very-high-energy photons (VHE, E$>$100 GeV) in active galactic nuclei (AGN) is closely connected with the production of ultra-relativistic particles. Among AGN, the subclass of extreme BL Lacertae are of particular interest because they challenge state-of-art models on how these cosmic particle accelerators operate. By cross-matching two gamma-ray catalogs (this is, 4FGL-DR2 and 2BIGB), we identified 23 high-synchrotron-peaked (HSP) blazar candidates with photometric or spectroscopic redshifts, good multi-wavelength coverage, that are possibly detectable by VHE instruments. We performed a new analysis of Fermi Large Area Telescope data including the effects of attenuation from the extragalactic background light and complemented these results by collecting multiwavelength data from optical, radio and X-ray archival observations. Their broadband spectral energy distributions were interpreted in terms of synchrotron-self-Compton models with external-Compton components and compared with the properties of prototypical extreme HSP blazars. Finally, we test their detectability with imaging atmospheric Cherenkov telescopes (IACTs) and propose a new method for selecting these extreme targets for these ground-based telescopes.}

\FullConference{37$^{\rm{th}}$ International Cosmic Ray Conference (ICRC 2021)\\
		July 12th -- 23rd, 2021\\
		Online -- Berlin, Germany}

\begin{document}
\maketitle

\section{Introduction}
%
%
%
%
The extra-galactic $\gamma$-ray sky is dominated by {\it blazars} \cite{4FGLpaper}, a peculiar class of radio-loud {\it Active Galactic Nuclei} (AGN) that host a collimated jet of relativistic plasma in close alignment with our line of sight. Due to the effects of relativistic beaming, the jet radiation dominates the source emission at almost every frequency and it produces a characteristic two-hump spectral energy distribution (SED), that extends from the radio domain all the way up to $\gamma$-ray energies. The low energy part of the SED, lying between radio and X-ray frequencies, is interpreted as Synchrotron radiation by relativistic charged particles in the magnetic field of the jet. The high energy part, which instead covers the $\gamma$-ray domain and may reach up to Very High Energies (VHE, $E > 100\,$GeV), is due to inverse Compton (IC) scattering of low energy photons over the emitting particles.

Depending on their optical properties, blazars are generally divided in {\it BL Lac} type objects, whose spectra are dominated by the synchrotron continuum and appear nearly featureless, showing at most weak emission or absorption lines, and {\it Flat Spectrum Radio Quasars} (FSRQ), which, instead, are characterised by prominent emission lines and a continuum with a relevant thermal contribution, suggestive of the presence of a radiatively efficient hot plasma accretion disk. In addition to the optical properties, another classification scheme is generally defined through the characteristics of the SED and, in particular, the position of the synchrotron peak frequency emission. In this way, blazars can be distinguished in Low Synchrotron Peaked (LSP, $\nu^S_{\mathrm{peak}} \leq 10^{14}\,$Hz), Intermediate Synchrotron Peaked (ISP, $10^{14}\, \mathrm{Hz} < \nu^S_{\mathrm{peak}} \leq 10^{15}\,$Hz) and High Synchrotron Peaked (HSP, $\nu^S_{\mathrm{peak}} > 10^{15}\,$Hz) sources. While LSP and ISP sources can be either BL Lacs or FSRQs, HSP objects are almost exclusively BL Lacs and are sometimes referred to as HBL. There is a growing evidence that a further class of extremely high synchrotron peaked BL Lac objects (EHSP, $\nu^S_{\mathrm{peak}} > 10^{17}\,$Hz) may exist. HSP and EHSP objects are of particular interest, because they represent the most frequent class of extra-galactic VHE sources.

Thanks to its all-sky monitoring observational strategy, the {\it Fermi} Large Area Telescope ({\it Fermi}-LAT \cite{LATpaper}) has detected $\gamma$-ray emission from thousands of objects that were subsequently identified as blazars. A fraction of these sources, however, have not yet been firmly classified, though showing blazar-like SED characteristics, and they are therefore referred to as blazar candidates of uncertain class (BCU). In this contribution we present the first results of a technique based on catalog matching and SED modelling that we developed to select EHSP source candidates from the {\it Fermi}-LAT BCUs. 

\section{Materials and Methods}
%
The sample analyzed in this study derives from a cross-match of the {\it Fermi}-LAT BCUs listed in the second release of the Fourth LAT AGN Catalog (4LAC-DR2, \cite{2020ApJ...892..105A}) with the Second Brazil-ICRANet Gamma-ray Blazars catalog (2BIGB, \cite{2020MNRAS.493.2438A}), which includes sources whose infrared photometric properties are similar to those of HBL sources. We limited our selection to objects that, according to the 2BIGB records, have a broad-band photometric coverage and for which we could obtain at least a redshift estimate (using either spectroscopic redshifts provided by \cite{2020ApJ...892..105A} and by \cite{2020arXiv201205176G}, or the 2BIGB's photometric estimates). We further introduced a {\it figure of merit} (FOM) parameter, defined as the ratio among the flux at synchrotron peak of any given source and the synchrotron peak flux of the faintest blazar presently detected at TeV energies, thereby considering only objects whose FOM is larger than 0.7. The result of this operation was the selection of 23 EHBL source candidates, listed in Table~1.

\begin{table}
\begin{Overpic}{
\begin{tabular}{lrrrrrrrr}
     4FGL Name &  RAJ2000 &  DEJ2000 &    $z$ &    TS & FOM & Index & VarIndex &      FracVar \\
\midrule
J0132.7$-$0804 &   23.183 &   -8.074 & 0.148  &    88 & 0.8 &   1.9 &     5.54 &  \\
J0212.2$-$0219 &   33.066 &   -2.319 & 0.250  &    61 & 0.8 &   2.2 &    16.70 &  $0.47\pm0.29$ \\
J0350.4$-$5144 &   57.613 &  -51.743 & $\sim$0.32  &    98 & 0.8 &   1.8 &    11.62 &  $0.32\pm0.36$ \\
J0515.5$-$0125 &   78.891 &   -1.419 & $\sim$0.25  &    55 & 0.8 &   2.1 &    13.04 &  $0.32\pm0.31$ \\
J0526.7$-$1519 &   81.692 &  -15.321 & $\sim$0.21  &   218 & 1.6 &   2.0 &     8.21 &  \\
J0529.1$+$0935 &   82.297 &    9.597 & $\sim$0.30  &    86 & 1.3 &   2.1 &    12.81 &  $0.23\pm0.26$ \\
J0557.3$-$0615 &   89.344 &   -6.265 & $\sim$0.29  &    53 & 1.6 &   2.0 &     7.08 &  \\
J0606.5$-$4730 &   91.642 &  -47.504 & 0.030  &   137 & 1.0 &   2.0 &    17.90 &  $0.39\pm0.20$ \\
J0647.0$-$5138 &  101.773 &  -51.638 & $\sim$0.22  &    81 & 2.5 &   1.8 &    17.03 &  $0.32\pm0.40$ \\
J0733.4$+$5152 &  113.362 &   51.880 & 0.065  &   162 & 2.5 &   1.8 &    12.43 &  $0.26\pm0.30$ \\
J0847.0$-$2336 &  131.757 &  -23.614 & 0.059  &   921 & 0.8 &   2.0 &    14.72 &  $0.14\pm0.09$ \\
J0953.4$-$7659 &  148.367 &  -76.993 & $\sim$0.25  &   104 & 0.8 &   2.0 &     5.88 &  \\
J0958.1$-$6753 &  149.534 &  -67.894 & $\sim$0.21  &    29 & 1.0 &   2.2 &    12.73 &  $0.46\pm0.51$ \\
J1132.2$-$4736 &  173.056 &  -47.613 & $\sim$0.21  &   129 & 1.0 &   2.0 &    11.47 &  $0.26\pm0.22$ \\
J1447.0$-$2657 &  221.765 &  -26.962 & $\sim$0.32  &    46 & 2.0 &   2.0 &     6.81 &  \\
J1714.0$-$2029 &  258.522 &  -20.486 & $\sim$0.09  &   110 & 2.0 &   1.6 &    24.95 &  $0.59\pm0.28$ \\
J1824.5$+$4311 &  276.126 &   43.196 & 0.487  &    99 & 0.8 &   1.8 &     8.17 &  \\
J1934.3$-$2419 &  293.582 &  -24.326 & $\sim$0.23  &    63 & 1.6 &   1.8 &    11.21 &  \\
J1944.4$-$4523 &  296.101 &  -45.393 & $\sim$0.21  &   164 & 1.0 &   1.7 &    14.32 &  $0.22\pm0.32$ \\
J2001.9$-$5737 &  300.491 &  -57.631 & $\sim$0.26  &   123 & 0.8 &   2.1 &     2.94 &  \\
J2142.4$+$3659 &  325.602 &   36.986 & $\sim$0.24  &   110 & 1.3 &   2.0 &    13.35 &  $0.25\pm0.29$ \\
J2246.7$-$5207 &  341.682 &  -52.126 & 0.194  &    95 & 2.5 &   1.7 &    19.10 &  $0.57\pm0.30$ \\
J2251.7$-$3208 &  342.944 &  -32.140 & 0.246  &    52 & 2.0 &   1.8 &    11.20 &  $0.28\pm0.49$ \\
\bottomrule
\end{tabular}}
\end{Overpic}
\caption{EHBL candidates considered in this work and main properties extracted from the 4FGL, 4LAC and 2BIGB catalogs.}
\end{table}

%
%
%
%

In order to optimize the coverage of the high energy emission, we repeated the $gamma$-ray analysis of the $Fermi$-LAT data using the same observation time interval of 4LAC-DR2, following the {\it Cicerone} analysis recommendations and taking into account the interaction between $\gamma$-ray photons and the Extragalactic Background Light radiation (EBL) at the modelling stage \cite{Dominguez15, Saldana20}. 
We subsequently built the multi-wavelength SEDs of our sources by searching for counterparts at different frequencies through the SSDC's SED builder tool.\footnote{\url{https://tools.ssdc.asi.it/SED/}} We looked for archival data covering from radio to X-rays that, together with our $\gamma$-ray spectra, resulted in a collection of SEDs that we modelled using the \texttt{jetset} emission code \cite{2020ascl.soft09001T}.

\begin{figure}
\centering
\includegraphics[width=0.75\linewidth]{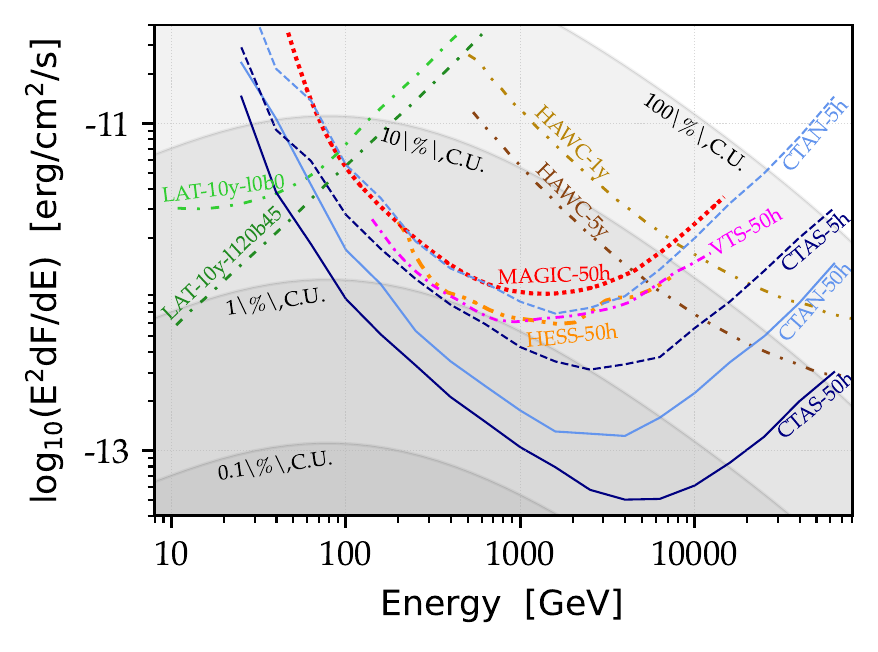}
\caption{Sensitivities of various $\gamma$-ray instruments in operation and planned.}
\label{fig:instrument_sensitivities}
\end{figure}

As a final step, we investigated the detectability of our sources in the VHE band by generating and anylising samples of simulated VHE $\gamma$-ray observations, based on the SED modeling previously introduced and using CTA's Prod3b response functions for an exposure time of $5\,$hr. Taking into account the estimated performance of currently existing and planned VHE observatories, some of which are illustrated as an example in Fig.~\ref{fig:instrument_sensitivities}, this approach is expected to approximately represent also the performance of currently operating Imaging Atmospheric Cherenkov Telescopes (IACT) in a cumulative exposure of $50\,$hr, thus illustrating the estimated detection possibilities of the selected targets with currently operating and future observing facilities.

\section{Properties of the selected BCU sample}
\begin{figure}
\centering
\hspace*{-7mm}\begin{overpic}[percent,width=0.55\linewidth]{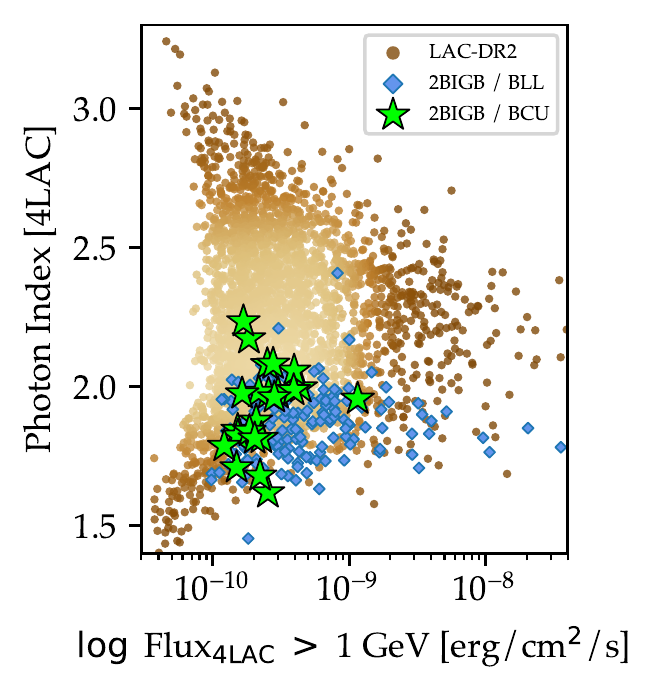}
\put(45,25){\color{red}\large Preliminary}
\end{overpic}
\hspace*{-9.5mm}\begin{overpic}[percent,width=0.5\linewidth]{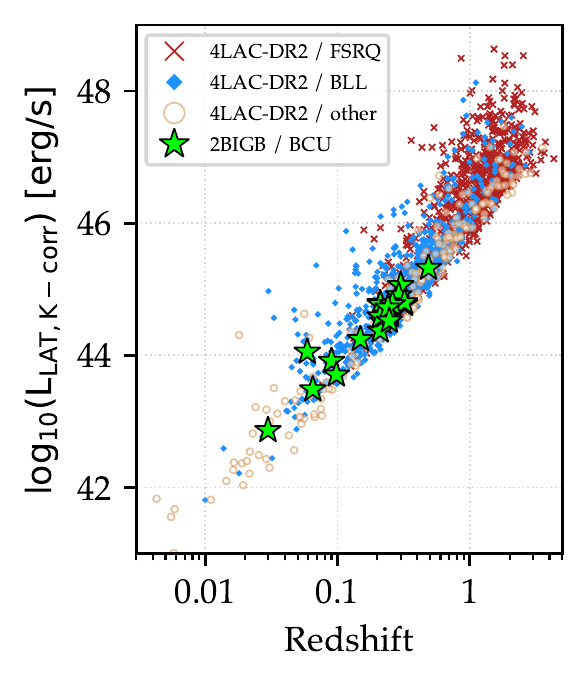}
\put(45,25){\color{red}\large Preliminary}
\end{overpic}
\caption{{\bf Left:} Spectral index vs. Flux in the Fermi-LAT band for: LAC-DR2 sources (density color coded brown tones), 2BIGB BL Lacs (blue) and our targets (green stars); {\bf right:} K-corrected $\gamma$-ray luminosity vs. redshift for FSRQs (red crosses), BL Lacs (blue  diamonds) and other extragalactic objects (open orange circles).}
\label{fig:index_vs_flux}
\vspace{1cm}
\end{figure}
%
In order to test the ability of our selection procedure to identify candidate HSP and EHSP sources, we compared the properties of the selected sources with those of already classified objects. Since we know that BL Lacs have statistically harder $\gamma$-ray spectra and are less luminous than FSRQs, we created diagnostic plots to illustrate the distribution of our sources with respect to the different blazar classes detected by {\em Fermi}-LAT. We can see from Fig.~\ref{fig:index_vs_flux} that our targets tend to have low to intermediate redshifts ($0.05 \lesssim z \lesssim 0.30$), hard spectra, like those of BL Lacs, but with rather low luminosities ($L_\gamma \lesssim 10^{45}\, \mathrm{erg/s}$).

\section{Broadband SED modelling}
A more precise identification of the most likely EHBL source candidates can be obtained by the construction of a set of SED models that reproduces the available observations. Most of the sources are well represented by the inclusion of some standard components. Specifically, these include:
\begin{description}
\item{\bf Non-thermal low-energy component:}
a one-zone leptonic model with emission originating in a spherical plasma region of radius $R$ embedded in the blazar jet. Its distance to the nucleus is $R_H$ and its magnetic field strength $B$. Electrons in the plasma are assumed to have a broken power-law spectrum with indices $p_1$ between Lorentz factors $\gamma_\mathrm{min}$ and $\gamma_\mathrm{br}$ and high-energy index $p_2$ between $\gamma_\mathrm{br}$ and $\gamma_\mathrm{max}$.
\item{\bf High energy emission:} Due to inverse Compton scattering from the same population of electrons on both synchrotron radiation and, if present, the infrared radiation from a dusty torus.
\item{\bf Thermal radiation:} A clear excess in the optical band is visible for all the sources, and assumed to be host emission from a giant elliptical galaxy. We model it as a black body with effective temperature $\mathrm{T_{eff,Host}}$ and luminosity $\mathrm{L_{Host}}$. For sources with a dusty torus, we assume a characteristic size $R_{DT}$, opacity $\tau_{DT}$ and effective temperature $T_{DT}$.
\end{description}

\begin{figure}\vspace{1cm}
\centering 
\begin{overpic}[percent,width=0.35\linewidth]{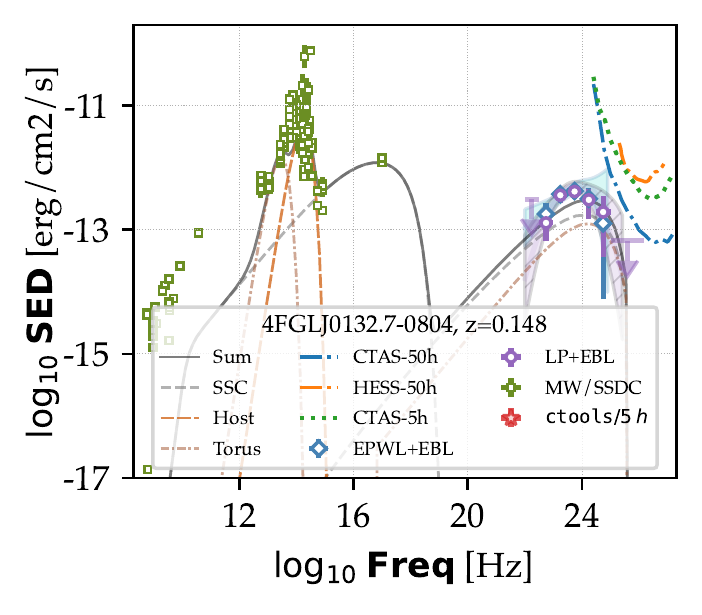}
\put(50,75){\color{red}\large Preliminary}
\end{overpic}
\begin{overpic}[percent,width=0.35\linewidth]{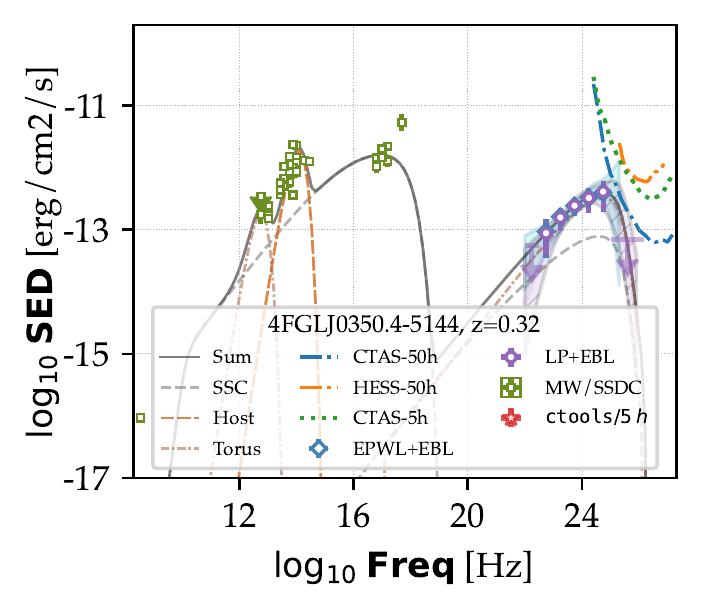}
\put(50,75){\color{red}\large Preliminary}
\end{overpic}
\begin{overpic}[percent,width=0.35\linewidth]{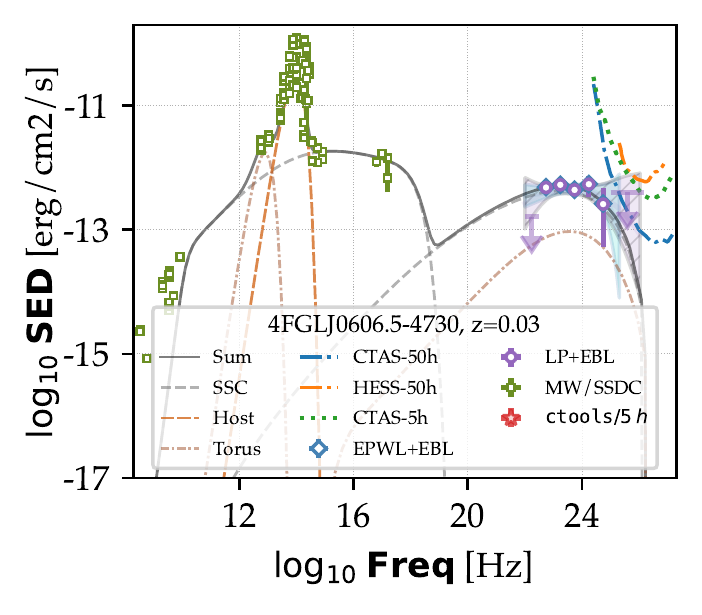}
\put(50,75){\color{red}\large Preliminary}
\end{overpic}
\begin{overpic}[percent,width=0.35\linewidth]{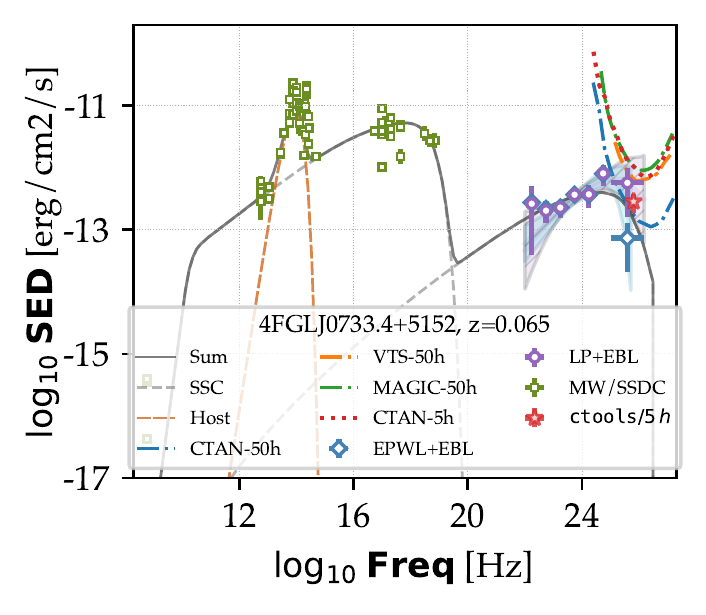}
\put(50,75){\color{red}\large Preliminary}
\end{overpic}
\begin{overpic}[percent,width=0.35\linewidth]{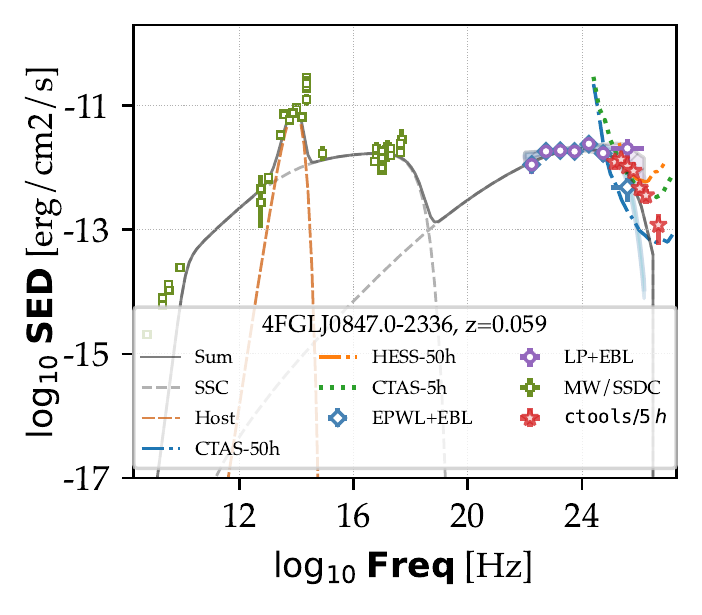}
\put(50,75){\color{red}\large Preliminary}
\end{overpic}
\begin{overpic}[percent,width=0.35\linewidth]{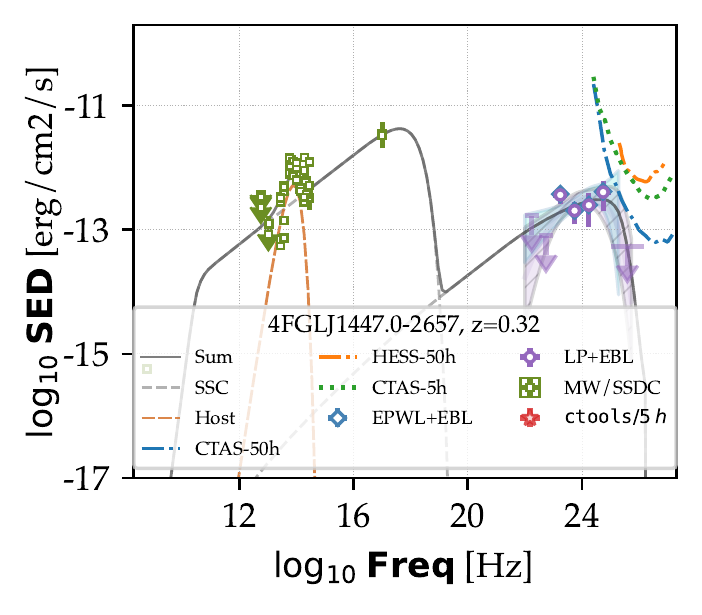}
\put(50,75){\color{red}\large Preliminary}
\end{overpic}
\begin{overpic}[percent,width=0.35\linewidth]{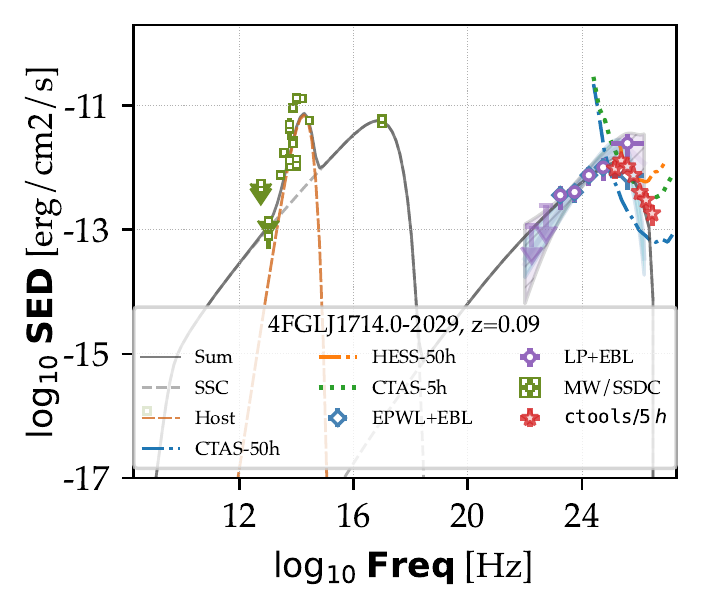}
\put(50,75){\color{red}\large Preliminary}
\end{overpic}
\begin{overpic}[percent,width=0.35\linewidth]{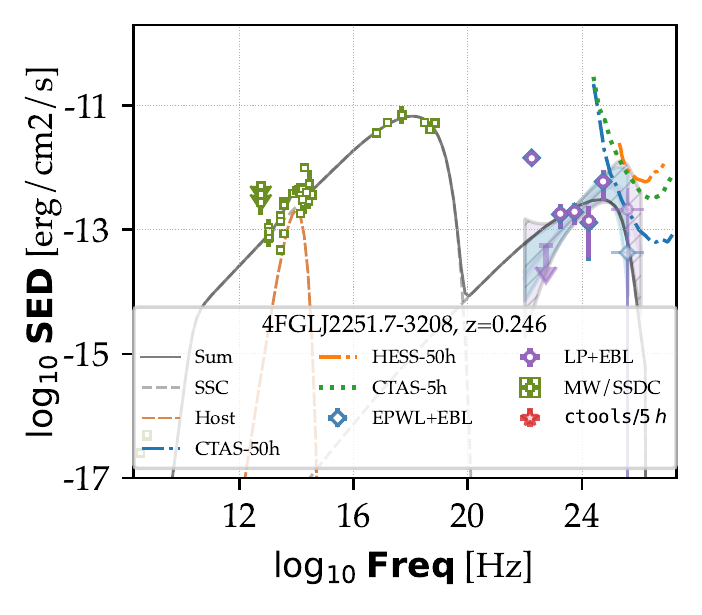}
\put(50,75){\color{red}\large Preliminary}
\end{overpic}
\caption{SEDs for 8 of the 23 sources considered in this work. Green open square markers represent archival data, purple and blue open circles show the Fermi-LAT analysis results (using a LogParabola and PowerLaw with exponential absorption shapes respectively, both absorbed by EBL) and red stars show the simulated $5\,h$ exposure observations with CTA, as a proxy for observations with existing IACTs in $50\,h$. Finally, the different curves show the different components of the resulting model.}
\label{fig:sed_models}
\end{figure}

The spectral fitting is calculated through the \texttt{jetset} code. Due the limitation of non-simultaneous input data and to the strong degeneracies existing between different properties, we fixed some parameters to a set of typical values. Specifically, we adopted a minimum particle Lorentz factor $\gamma_{min}= 1$, a bulk jet Lorentz factor $\Gamma = 20$, a scale radius of the emitting region $R = 1\times 10^{16}\,\mathrm{cm}$, and a distance of the emitting region from the center $R_H=2\times10^{18}\,\mathrm{cm}$. 
%
%
The resulting SEDs for 8 sources out of the 23 included in our sample are shown as an example in Fig.~\ref{fig:sed_models}.

\section{Conclusions}
In this study, we explored the possibility of using a cross-match of catalogs of photometric data to identify potential HSP and EHSP emitters within the sample of {\it Fermi}-LAT BCU sources and to reconstruct their SEDs. Subsequent modelling of their broadband emission confirmed that 16 objects, out of the original sample of 23 sources, can be classified as EHSP, while the remaining ones have an HSP SED. Interestingly, we note that one-zone leptonic models are able to provide good fit to all the spectra. We also observe thermal components, most frequently as a strong host galaxy contribution, which is commonly expected in EHSP, but with hints of unexpected torus-like IR excesses in few cases. Three sources, namely 4FGL J0733.4+5152, 4FGL J1447.0-2657 and 4FGL J2251.7-3208m exhibit extreme synchrotron peak frequencies ($\nu^S_{\mathrm{peak}} > 10^{18}\,$Hz). Only the first and the last have enough X-ray data to clearly constrain the peak position, while for 4FGL J1447.0-2657 it was constrained mainly from the Fermi-LAT spectrum (therefore the peak frequency should be considered a lower limit). The VHE visibility tests resulted in the identification of only 4FGL J0847.0-2336 and 4FGL J1714.0-2029 as promising VHE candidates, while the remaining sources have too low predicted fluxes or soft spectra to be considered possibly interesting targets. Both these sources have, according to the proposed radiative model, very low magnetization ($B\sim10^{-4}\,\mathrm{G}$), strong host emission and no IR torus.
%

Based on these results, we conclude that the SED of HSP candidates can be well reproduced within a Synchrotron-Self Compton scenario (SSC), based on a rather standard set of spectral components, even with the limitations imposed by non simultaneous observations. Clearly, the availability of more data, particularly in the X-ray domain, would help us to better constraint the physical parameters of the models, but their ability to reproduce the overall appearance of the SED provides an extremely useful tool to test the HSP nature candidate sources identified by means of multi-wavelength catalog searches.

\section*{Acknowledgements}
The \textit{Fermi}-LAT Collaboration acknowledges support for LAT development, operation and data analysis from NASA and DOE (United States), CEA/Irfu and IN2P3/CNRS (France), ASI and INFN (Italy), MEXT, KEK, and JAXA (Japan), and the K.A.~Wallenberg Foundation, the Swedish Research Council and the National Space Board (Sweden). Science analysis support in the operations phase from INAF (Italy) and CNES (France) is also gratefully acknowledged. This work performed in part under DOE Contract DE-AC02-76SF00515.

\nocite{*} 

\bibliographystyle{unsrt}
\bibliography{sample} 

%
%
%

\end{document}